\documentclass[a4paper]{jpconf}
\usepackage{graphicx}

\newcommand{\cm}{\; \mathrm{cm}}
\newcommand{\GeV}{\; \mathrm{GeV}}

\begin{document}
\title{Neutrino nucleus reactions within the GiBUU model}

\author{O. Lalakulich, K. Gallmeister, U. Mosel}

\address{Institut f\"ur Theoretische Physik, Universit\"at Giessen, Germany}

\ead{Olga.Lalakulich@theo.physik.uni-giessen.de}

\begin{abstract}
The GiBUU model, which implements all reaction channels relevant at medium neutrino energy,
is used to investigate the neutrino and 
antineutrino scattering on iron. Results for integrated cross sections are compared with NOMAD and MINOS data.
It is shown, that final state interaction can noticeably change the spectra of the outgoing hadrons.
Predictions for the Miner$\nu$a experiment are made for pion spectra, averaged over 
NuMI neutrino and antineutrino fluxes.

Contribution to NUFACT 11, XIIIth International Workshop on Neutrino Factories, 
Super beams and Beta beams, 1-6 August 2011, CERN and University of Geneva
\end{abstract}

\section{Introduction}

Neutrino and antineutrino scattering on nuclei for neutrino energies above $30 \GeV$ was studied 
in several experiments starting from the 80s. Theoretically they were successfully described within
the quark parton model as Deep Inelastic Scattering (DIS) processes.
Recent measurements by MINOS and NOMAD colalborations also
covered the intermediate energy region. Here the neutrino reactions are not so easy to model, because of
the overlapping contributions from QE scattering, resonance production and 
background processes.
This requires complex approaches that take all of the relevant channels into account.

Nuclear effects in neutrino reactions can also be studied in detail nowadays. 
The Miner$\nu$a experiment intends to perform measurements on
Plastic (CH), Iron, Lead, Carbon, Water and liquid Helium targets in the NuMI beam, 
which would directly  allow to compare nuclear effects on various nuclei.  
Besides muon detection, this experiment will also be able to resolve various final states
by identifying the tracks of the outgoing hadrons. Theoretical modelling of such exclusive reactions
requires realistic approaches to the  desctiption of the initial and final state interactions in target nuclei.

Both these requirements are satisfied by the GiBUU transport model \cite{gibuu,Buss:2011mx}, which we use here 
to study neutrino and antineutrino scattering on iron. Our results  are compared
with the recent MINOS and NOMAD data.  Predictions are also made for the spectra of the outgoing particles.
The calculations are done without any fine tuning to the 
data covered here with the default parameters as used in the GiBUU framework.

\section{GiBUU transport model}

The GiBUU model was initially developed 
as a transport model for nucleon-, \mbox{nucleus-,}  pion-, and electron- induced reactions from
some hundreds MeV up to tens of GeV. Several years ago neutrino-induced interactions were
also implemented for the energies up to a few GeV. 
Recently the code was extended to describe also the DIS processes in neutrino reactions.
Thus, we can study all kind of elementary collisions on all kind of nuclei within a unified framework. 
The model is based on well-founded theoretical ingredients and has been tested against various nuclear reactions.
For a detailed review of the GiBUU model see \cite{Buss:2011mx}.

GiBUU describes all processes relevant at medium energies, the cross section is calculated as
$\sigma_\textrm{tot}=\sigma_{QE}+\sigma_{RES}+\sigma_{BG}+\sigma_{DIS}$.
Our approach to quasielastic (QE) scattering, resonance (RES) production and background (BG) processes 
is described in \cite{Leitner:2006ww,Leitner:2008ue}. 
The DIS scattering is  included   as \textsc{Pythia} simulation.

In the region of the shallow inelastic scattering, that is at moderate 
invariant masses, $1.6\GeV<W<2.0\GeV$,  there is a potential problem of double counting.  
Here the same physical  events can be considered as originating from decays of 
high mass baryonic resonances  or from DIS.
In the GiBUU code we use the ansatz, that both RES, BG and DIS processes contribute in this region.
While the RES and BG contributions are smoothly switched off
in this region and DIS contribution is switched on.
With this choice, the DIS events become noticeable at neutrino energies around $3\GeV$.  
In essence, the DIS processes below at lower $W$ account for the
resonances whose electromagnetic properties are not known  and for the 
non-resonant processes giving several mesons in the final state beyond the one pion background.

\section{Integrated cross sections.}

As was measured in 80s, the DIS cross section grows linearly with energy. 
So at high neutrino energies the data are conveniently presented as cross section  per energy $\sigma_\textrm{tot}/E_\nu$.  
Despite the measurements being made on nuclear targets,
the world average values are given \cite{Amsler:2008zzb} for isoscalar target:
$\sigma_{tot}/E_\nu (\nu) = 0.667\pm 0.014 \cdot 10^{-38} \cm^2/\GeV$ for neutrinos
and $\sigma_{tot}/E_\nu (\bar\nu) = 0.334\pm 0.008 \cdot 10^{-38} \cm^2/\GeV$ for antineutrinos.
Such an approach is meaningful, only if nuclear corrections are neglected.

\begin{figure}[hbt]
\begin{minipage}[c]{0.48\textwidth}
\includegraphics[width=\textwidth]{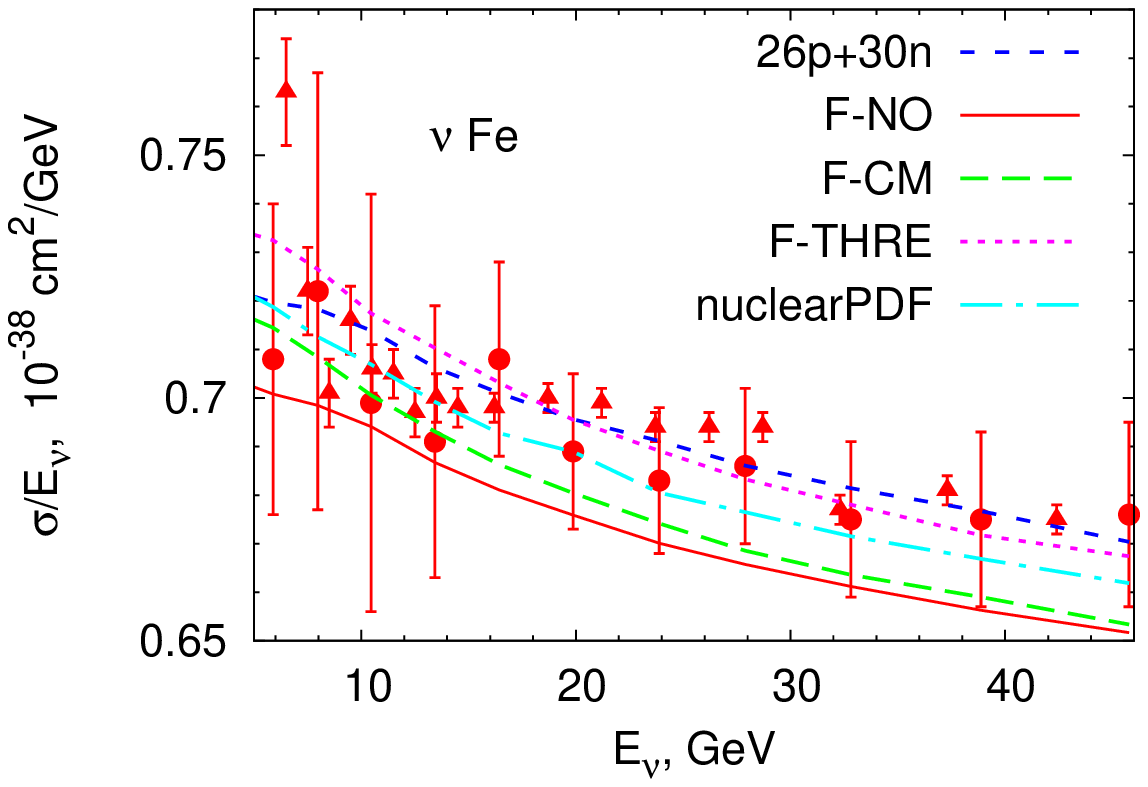}
\end{minipage}
\hfill
\begin{minipage}[c]{0.48\textwidth}
\includegraphics[width=\textwidth]{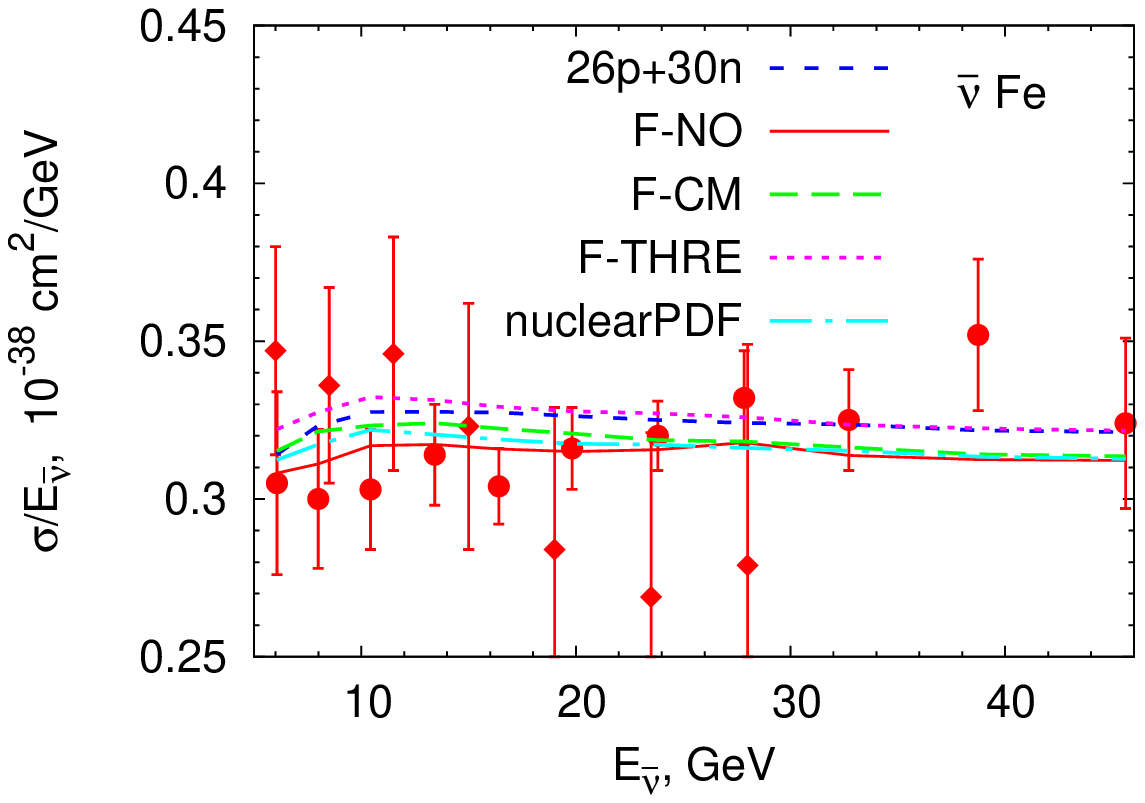}
\end{minipage}
\caption{(Color online) Total cross section per nucleon for neutrino (left) and antineutrino (right) 
induced reactions on iron target. Various prescriptions to include nuclear effects in DIS are compared.
MINOS data (full circles) are from \protect\cite{:2007rv,Adamson:2009ju}, 
IHEP-JINR data (full diamonds) are from \protect\cite{Anikeev:1995dj}.}
\label{fig:compare-nuclear}
\end{figure}

The actual value of such corrections for neutrino reactions is not known so far, 
because of both experimental inaccuracies and difficulties in the theoretical description. 
On one hand, nuclear parton distributions, based on electromagnetic scattering data and intended for description
of both charged lepton and neutrino reactions, were introduced. For a review and a list 
of recent parametrization see, for example, \cite{Hirai:2009mq}.
On the other hand,
recent investigation \cite{Schienbein:2007fs,Kovarik:2010uv} showed, that in neutrino reactions
nuclear corrections to parton distributions are at the same level as for electrons, but
have a very different dependence on the Bjorken $x$ variable.
The topic remains controversial, with the hope that future precise Miner$\nu$a results on various
targets will clarify the situation. 

As we already mentioned, the GiBUU code uses PYTHIA for the simulation of DIS processes.
In the GiBUU simulation the neutrino interacts with one initial nucleon, bound in the hadronic potential and having nonzero
Fermi momentum. In order to be able to use the PYTHIA event simulator, we have to provide some
quasi-free kinematics as inputs to PYTHIA. Various prescriptions to do this 
result in a $5-7\%$ difference in the results. 
The corresponding cross sections (denoted as ``F-NO'', ``F-CM'', ``F-THRE'') are shown in Fig~\ref{fig:compare-nuclear}.
Nuclear parton distribution functions
from \cite{Eskola:1998df} are also implemented as one of the options to use 
(to avoid double counting, nuclear potential and Fermi motion in such calculations are switched off). 
The result (``nuclearPDF'') as well as the free cross section for iron composition 
(``26p+30n'') are also shown in Fig~\ref{fig:compare-nuclear}. 
At the moment we consider the various prescriptions mentioned above as intrinsic uncertainty 
of the GiBUU code, reflecting the lack 
of our understanding the nuclear effects. 
No other event generator, as far as we know, accounts for nuclear corrections in high--energy neutrino reactions.  

Fig.~\ref{fig:compare-nuclear} shows, that our calculations are in a good agreement with the recent neutrino data,
which are also consistent with each other.  
This figure shows that the decreasing slope of our curves for the neutrino cross section
is in agreement with that of the data. This slope is not taken into account in deriving the world-average
value, where it was assumed to be negligible. 
For antineutrino the agreement is good for $E_{\bar\nu}>25\GeV$.
For lower energies our curve is  above the recent MINOS
data, but below the IHEP-JINR results \cite{Anikeev:1995dj}. The overall agreement of our calculations with the data is
therefore better than the agreement of the data with each other.

\section{Final state interactions and change of the final hadronic spectra.}

After being produced in the initial interaction, outgoing hadrons propagate throughout the nucleus.
In GiBUU this process of final state interactions (FSI) is modeled by solving the semi-classical 
Boltzmann-Uehling-Uhlenbeck equation.
It describes the dynamical evolution of the phase space density for each particle species
under the influence of the mean field potential, introduced in the description of
the initial nucleus state. Equations for various particle species are coupled through this mean field and
also through the collision term. This term explicitly accounts for changes in
the phase space density caused by elastic and inelastic collisions between particles.
FSI decrease the cross sections as well as significantly modify
the shapes of the final particle spectra. Such change was seen, for example,
in photo-pion production~\cite{Krusche:2004uw} and is described by the GiBUU with a good accuracy.
A similar change should be observed in neutrino reactions.

\begin{figure}[hbt]
\centering
\includegraphics[width=0.8\textwidth]{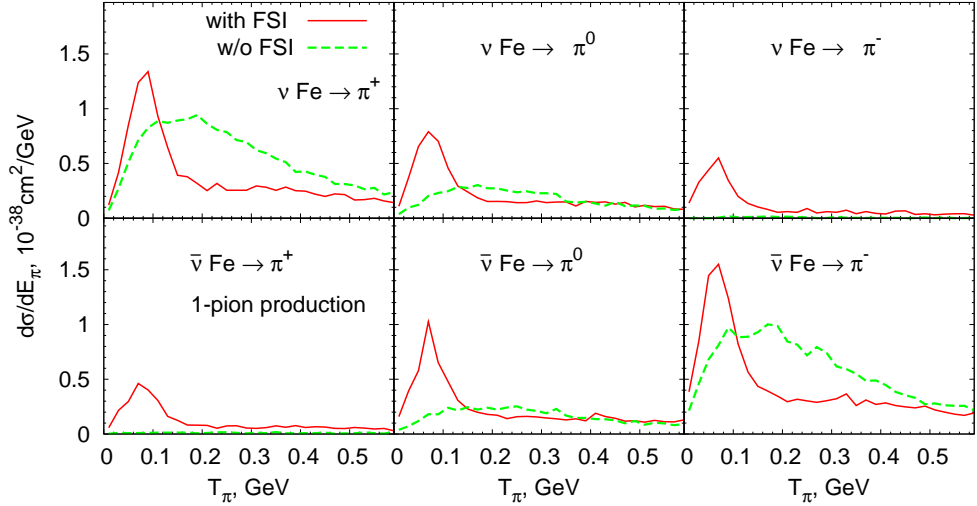}
\caption{(Color online) Pion kinetic energy distributions for neutrino and antineutrino induced reactions for 1-pion production
(one pion of a given charge and no other pions are produced).  Calculations are for the NuMI low-energy mode neutrino/antineutrino fluxes.}
\label{fig:MINOS-ekin-pion-with-FSI-1-pion}
\end{figure}

Fig.~\ref{fig:MINOS-ekin-pion-with-FSI-1-pion} shows the $\pi^+$ (left panels), $\pi^0$ (middle panels)
and $\pi^-$ (right panels)  spectra for neutrino (upper panels) and antineutrino (lower panels) NuMI
fluxes for 1-pion events.  The green dashed lines show the kinetic energy ($T_\pi$) distributions without FSI,
i.e. of pions produced in the initial neutrino vertex. The solid red lines show the distributions
after FSI, i.e.  of pions that made it out of the nucleus.
Such spectra can also be calculated for any other outgoing particles (protons, kaons, eta)
for any predefined final state and should be measurable in Miner$\nu$a experiment.

For dominant channels ($\pi^+$ production for neutrino reactions and $\pi^-$ in antineutrino ones),
the FSI decrease the cross section at $T_\pi > 0.2 \GeV$. This is mainly explained by pion 
absorption through $\pi N \to \Delta$ following by $\Delta N \to NN$. 
Pion elastic scattering in the FSI also decreases the pion energy,
thus depleting the spectra at higher energies and accumulating strength at lower
energies. Thus, an increase of the cross sections is observed at $T_\pi < 0.15 \GeV$;
altogether this leads to a significant change of the shape of the spectra.

Scattering can also lead to pion charge exchange. For neutrino-induced reactions, 
the  $\pi^+ n \to \pi^0 p$ scattering in the FSI is the main source of side--feeding 
for the $\pi^0$ channel, leading to a noticeable increase of the 
$\pi^0$ cross section at low $T_\pi$.  The inverse feeding is suppressed, because less 
$\pi^0$ than $\pi^+$ are produced at the initial vertex. The same mechanism of side feeding
from dominant to sub-dominant channel through $\pi^- p \to \pi^0 n$
is working for antineutrino induced reactions.

For the least dominant channel ($\pi^-$ production for neutrino reactions and $\pi^+$ in antineutrino ones),
the FSI (in particularly side feeding)  represent the main source of the events observed; thus a dramatic 
FSI effect.

\section*{Acknowledment}

This work is supported by DFG. O.L. is grateful to Ivan Lappo-Danilevski for programming assistance.

\bibliographystyle{iopart-num}
\bibliography{nuclear}

\end{document}